# Effect of rotation of feed and seed rods on the quality of $Na_{0.75}CoO_2$ single crystal grown by traveling solvent floating zone method

C. Sekar[a,*], S. Paulraj[b], P. Kanchana[b], B. Schüpp-Niewa[c], R. Klingeler[c, d], G. Krabbes[c], B.Büchner[c]

[a]*Department of Bioelectronics & Biosensors, Alagappa University, Karaikudi 630 003, India*

[b]*Department of Physics, Periyar University, Salem 636 011, India*

[c]*Leibniz-Institute for Solid State and Materials Research, IFW-Dresden, 01171 Dresden, Germany*

[d]*Kirchhoff Institute for Physics, University of Heidelberg, D-69120 Heidelberg*

**Abstract**

High purity $Na_{0.75}CoO_2$ single crystals have been grown by floating zone method. We found the rotation of feed and seed rods play a crucial role in growing high quality single crystal. Systematic investigations suggest the occurrence of a phase separation at microscopic level, such as the separation into Na-rich and Na-poor domains during the growth, and formation of impurity phase(s) depending on growth conditions. $Na_xCoO_2$ ($x$ = 0.30, 0.60) crystals have been prepared by sodium deintercalation from $Na_{0.75}CoO_2$. Powder X-ray and energy dispersive X-ray analyses have confirmed the phase purity and homogeneity of the samples. Magnetic susceptibility measurements of $x$ = 0.60 and 0.75 crystals indicate a bulk phase transition at 22 K and an anomaly around 339 K and 334 K respectively which are attributed to structural transition. In addition some unique features related to $Na_{0.60}CoO_2$ were observed.

*Corresponding author. Tel: +91 4565 226385; Fax: +91 4565 225202
*E-mail*:Sekar2025@gmail.com

## 1. Introduction

Sodium cobalt oxide ($Na_xCoO_2$) has attracted considerable attention because of large thermoelectric power coupled with metallic resistivity in $Na_{0.50}CoO_2$ single crystals [1] and the discovery of superconductivity below 5 K in a hydrated version of this material ($Na_{0.35}CoO_2 \cdot 1.3 H_2O$) [2]. Furthermore, the structural and chemical similarities between $Na_xCoO_2$ compounds and cuprate superconductors have rekindled the hope of additional insight into the physics of superconductivity in layered transition metal oxides. The basic structure of $Na_xCoO_2$ ($0.3 < x < 0.9$) consists of planes of edge-sharing $CoO_6$ octahedra alternating with layers of Na. The γ-$Na_{0.7}CoO_2$ has hexagonal ($P6_3/mmc$) structure with $a = 2.83$ Å, and $c = 10.85$ Å. The ability to control the sodium content is an important requirement as it has the influence on the $Co^{3+}/Co^{4+}$ ratio and the physical properties [3].

Several methods have been reported to prepare polycrystalline sample with a desired composition in the $Na_xCoO_2$ system [4-5]. Single crystals of $Na_{0.75}CoO_2$ have been grown by flux method [6-8] and floating zone (FZ) method [9-14]. Sales et al. [9] reported the first FZ growth of $Na_{0.75}CoO_2$. Subsequently, different approaches have been proposed in order to improve the crystal size and quality. Prabhakaran et al. [12] reported that the crystal growth under high-pressure atmosphere minimizes Na loss significantly which suppresses the formation of $Co_3O_4$ impurity phase in the grown crystal. Peng and Lin [13] reported that the nominal growth rate of 2 mm/h under flowing oxygen atmosphere is the optimum condition to grow crystals of $Na_{0.75}CoO_2$. Fast pulling rate (more than 8 mm/h) in oxygen atmosphere were reported to yield large crystals with lesser CoO impurity and the crystals grown in this way were reported to be easy to cleave [14]. In general, single crystals grown under slightly different conditions are reported to contain different types of chemical impurities (e.g. CoO, $Co_3O_4$, $Na_2O$ and Na poor $Na_xCoO_2$ phases) and exhibit qualitatively different properties. There are two major difficulties in the crystal growth of $Na_xCoO_2$; (i) the loss of Na during the growth due to its volatile nature and (ii) phase inhomogeneity in the $Na_xCoO_2$ system due to the existence of wide range of solid solution.

In the present work, we have carried out a systematic investigation on the influence of growth parameters on crystal quality, phase homogeneity, and impurity formation. The work is focused particularly on the studies concerning relationship between starting composition ($x$ = 0.75 and 0.80) and growth parameters (mainly rotation of feed rod and growing crystal in opposite directions) on the crystal quality. The segregation coefficient of Na in the grown crystals with $x$ = 0.75 and 0.80 have been estimated in order to assess its distribution in the crystals.

## 2. Experimental

*2.1 Synthesis*

Powder samples of $Na_xCoO_2$ ($x$ = 0.75, 0.80 and 0.90) were prepared via solid-state reactions of $Na_2CO_3$, and $Co_3O_4$ with Na to Co ratio of 0.75-0.90. In all the three cases 5 wt. % $Na_2CO_3$ was added to the stoichiometric mixture in order to compensate the Na loss during the synthesis. The thoroughly mixed and ground powder was calcined at 750ºC for 12 h and then reacted at 850ºC for 2×16 hrs with intermittent grindings. After grinding thoroughly, cylindrical rods (~ 6 mm diameter & 100 mm long) were made by hydrostatically pressing under 15 kN/cm². The polycrystalline rod was heat treated at 850ºC for 10 h, and further densification at 900ºC for 6 h in flowing oxygen atmosphere. In the latter case, the rod was set to both translational and rotational motion in order to make it more homogeneous. Crystal growth was carried out by FZ method in an infrared radiation furnace equipped with four ellipsoidal mirrors (Crystal Systems

Inc.). The feed rod and the growing crystal were rotated in opposite directions. Following variable growth parameters were chosen in order to find optimum conditions; (i) growth rate 2-5 mm/h, (ii) growth atmosphere (flowing $O_2$: 20, 30, 50, 100 ml/min and high oxygen pressure of ∼ 8 bar and (iii) rotation rate (15, 20, 25 and 30 rpm).

*2.2 Characterization*

Thermogravimetry-Differential thermal analysis (TG-DTA) was carried out by measuring weight loss and calorimetric signals from room temperature to 1200ºC with a heating rate of 10 K/min in oxygen atmosphere using TG-DTA 92 SETARAM thermoanalyser. Phase purity of all the samples was checked by powder XRD measurements on polycrystalline samples and crushed single crystals. X-ray diffraction patterns were recorded (PHILIPS PW1820) using Co$K_\alpha$ radiation and a secondary graphite monochromator. The chemical compositions of the grown crystals were determined with a Philips XL 30 using back-scattered scanning electron microscope (SEM) and energy dispersive analysis of X-rays (EDX). Different samples were probed at various spots and the average value of the element concentration was taken to determine the final composition. In bulk samples, the Na content was estimated by inductively-coupled plasma mass spectroscopy (ICP-OES) analysis using Element XR-UV Microprobe II laser system (VG Elemental). The dc magnetic susceptibility χ of all compounds was determined from χ = $M/H$ in the temperature range from 2 to 360 K, where the magnetization $M$ was measured in a Quantum Design SQUID magnetometer at an applied field of 1 T.

**3. Results and Discussion**

*3.1 Thermal analysis*

In order to understand the influence of starting composition on the melt behavior, TG-DTA measurements have been carried out on $Na_xCoO_2$ ($x$ = 0.75, 0.80 and 0.90) powder samples.

Fig.1 shows the TG-DTA results for $x = 0.75$ sample. The sample melts incongruently near 1035°C ($T2$) and subsequently enters into a complete liquid phase near 1055±5°C. The peaks on DTA curve correlate well with the weight loss at the corresponding temperatures as revealed by TG data. The $x = 0.80$ and 0.90 samples also exhibited similar thermal behavior. Close observations of the TG results show that the decomposition temperatures decrease with increase in Na content. An additional low temperature event at around 850°C ($T1$) was present in $x = 0.80$ and 0.90 samples. This reaction may correspond to the decomposition of unreacted $Na_2CO_3$ or NaOH which are hard to detect by powder XRD measurement. In addition, a significant increase in mass loss has been observed in these samples when compared to that of $x = 0.75$. Table.1 shows the decomposition temperatures and mass loss of $Na_xCoO_2$ ($x = 0.75$, 0.80 and 0.90). Solidification (exothermic) process of all the investigated $Na_xCoO_2$ samples ($x = 0.75$, 0.80 and 0.90) shows one common broad peak at 1031°C (peak temperature) suggesting that one particular composition or its solid solution crystallizes in all cases irrespective of the starting composition. Other secondary phase(s) solidify during further cooling of the melt.

*3.2 Crystal growth*

In general, several growth factors such as pulling rate, the rotation rate of the seed and feed rods in opposite direction to make the homogeneous melt, the growth atmosphere and its pressure have to be optimized to grow high quality crystals. In the present work we have optimized these conditions to grow good quality single crystals of $Na_{0.75}CoO_2$.

*3.2.1 Crystal growth of $Na_{0.75}CoO_2$*

Initially $Na_{0.75}CoO_2$ crystals were grown under flowing oxygen atmosphere (30 ml/min) with a pulling rate of 2 mm/h using densified polycrystalline rod as a seed. Both seed and feed rods were rotated at 20 rpm in opposite directions. After a few hours of growth, the melt zone became unstable. Hence the feed rod was supplied rapidly in order not to disrupt the melt zone. The melt got accumulated and fell down resulting in the formation of a bunch of solid around the tip of the grown crystal. Subsequently, stable melt zone was formed on the tip of the grown crystal (seed) and the growth continued till the end. We assumed that the melt zone instability during initial stages of growth might have caused by porous nature of the seed rod due to melt sucking. With further insertion of feed rod, melt zone composition might have self adjusted enabling the stable growth.

Then the next growth experiments were performed under similar growth conditions using previously grown crystal rod as seed. Here, the melt zone was stable throughout the growth experiment which yielded about 80 mm long and 6 mm diameter crystal boule. The total mass loss during the growth was estimated as 2.4 wt. % by comparing the total mass of the products before and after the crystal growth. Several slices were made by cutting the rod perpendicular to its growth direction. SEM pictures of the cut and polished crystals (Fig. 2) revealed the presence of impurity phase which was identified as $CoO_{1-\delta}$. Though the oxygen content could not be estimated precisely using EDX, there has been a systematic and clear indication that these tiny impurity crystals are oxygen deficient.

Crystals grown at higher growth rates (> 3.0 mm/h) revealed cellular interface that is an indication for unstable growth. Intergrowth of other compositions and impurity phases were typically found at the cellular interface under such high growth rate conditions. Hence, further experiments were aimed at optimizing other growth parameters such as growth atmosphere and rotation rates under the fixed growth rate of 2.5 mm/h.

Growth experiments made under enhanced oxygen flow rate (100 ml/min) showed a mass loss of 3.8 wt. % which is slightly higher when compared to 2.4 wt. % for 30 ml/min. Crystal boules grown in this way appeared more shiny, but the inner core of the boule still contained inclusions of impurity phase(s). These impurity phase(s) are concentrated mostly at the inner core of specimens and there was a significant variation in the Na concentration along the radial and longitudinal direction of the crystal. These impurities were identified as $Co_3O_4$ with nearly optimum stoichiometry. This result is in accordance with the fact that the phase stabilization of either $Co_3O_4$ or $CoO$ depends essentially on the atmosphere.

Similarly growth experiments under high pressure oxygen (up to 8 bar) neither improved the phase homogeneity nor suppressed the impurity formation completely. The typical mass loss during the growth is about 2 wt. % of the total mass. A white colored deposit ($Na_2O$) was noticed on the inner wall of the quartz tube as in the case of growth under ambient oxygen atmosphere. High oxygen pressure does not seem to suppress the loss of Na vapor significantly.

*3.2.2 Optimal rotation rate*

It is known that the optimum rotation rate can usually lead to highly homogeneous melt with uniform distribution of solute particles, thus avoiding the solidification of undesired phase(s). Various rotation rates (15, 20, 25 and 30 rpm) of the feed rod and the seed rod were chosen to induce alternate magnitudes of forced convection, thereby altering the shape and curvature of the growth interface. Figs. 3a-d show the cross sectional view of the $Na_{0.75}CoO_2$ crystals grown at different rotation rates by keeping other parameters constant (growth rate 2.5 mm/h and oxygen flow rate 30 ml/min). As can be seen in the pictures, the shape and amount of the impurities decrease with increase in rotation rate. The corresponding impurities were identified as $Co_3O_4$, $CoO$ and $Na_{0.15}CoO_2$. At fixed growth rate (2.5 mm/h), the melt zone was stable under the rotations of 20-30 rpm. However, it was not possible to grow long crystal boules at higher rotation rates (> 30 rpm). The melt zone got interrupted suddenly after about 20-30 mm long rods were grown. It may be that the pulling rate has to be increased in order to complete the growth of the long crystal boule. However, higher growth rate causes the formation of cellular interface comprising other (Na/Co) compositions and impurity phases. Contrary to this, long crystal boules could be grown with lower rotation (15 and 20 rpm) rate but they also have a significantly large variation in Na content and impurity phases. The results suggest that the optimum rotation rate may be 25 rpm in flowing oxygen atmosphere (30 ml/min) and at the growth rate of 2.5 mm/h. The typical crystal boule contained a large single crystal of several mm thick with the large faces parallel to the growth direction. The $c$ axis is always perpendicular to the growth direction and the crystal is relatively easy to cleave along the $ab$ plane. EDX results indicate the Na content to be near 0.75 ($\pm$ 0.03) at different parts of the crystal and that there is no significant difference in the Na content between the core and outer ring regions of the crystal boule. Figs.4a and b show the SEM pictures of $Na_{0.75}CoO_2$ crystal taken along the horizontal and vertical cross-section respectively.

*3.2.3 Crystal growth of $Na_{0.80}CoO_2$*

Large crystal boules have been successfully grown from Na-rich composition ($x = 0.80$) following the conditions optimized for $Na_{0.75}CoO_2$. Detailed SEM-EDX analyses revealed that the average Na content in the grown crystals do not agree with the nominal composition and that the inner core of the cylindrical rods always had Na-poor composition ($x = 0.70$) while the outer core of the rod had the desired composition ($x = 0.80$). This is possibly because of the higher Na content in the melt and temperature gradient in the melt zone along radial cross-section. There were inclusions of impurity phase ($Co_3O_4$) and quantity of these tiny crystallites decreased along the longitudinal direction of the crystal boule. Fig. 5 shows the presence of impurities in the middle portion of the crystal boule grown using $Na_{0.80}CoO_2$ compound as feed. However, these impurities were minimal in tail of the boule and high purity crystals with dimensions of about 8×2 mm$^2$ could be cleaved from the last portion of the boule.

*3.3 Segregation coefficient*

In order to grow solid solution single crystals by FZ method with a fixed composition, it is essential to know the segregation coefficient (*K*) of the admixture in the host lattice. The ratio of concentration of a given atomic species in the solid phase ($C_s$) to that in the liquid phase ($C_l$) at equilibrium is called the segregation coefficient for that element and is denoted by $K = C_s/C_l$. In FZ method, axial segregation of solute occurs due either to convection or to forced flow driven, for example, by crystal rotation. The degree of that axial segregation depends on the extent to which the solute segregation coefficient (*K*) differs from unity.

Here, we have estimated the Na content measured on freshly cleaved crystals ($C_l$) by EDX method and the nominal Na content in the feed rod ($C_s$) by ICP-OES method. Fig. 6 shows the segregation coefficient (*K*) as a function of length of the crystal boule grown under identical conditions. The segregation coefficient *K* was found to be slightly lower than unity in the bottom portion of the *x* = 0.75 boule which indicate that the actual Na content in the melt is higher than the starting composition. Subsequently, the melt self adjusted by itself leading to near optimum growth. Further growth resulted in the Na-poor crystal (*K*>1) in the tail part of the boule. This could be due to the loss of Na due to its volatile nature. The crystals obtained from the mid portion till the tail of the boule were found to be clean and have nearly uniform Na content across the radial cross-section. On the other hand, the segregation coefficient (*K*) of *x* = 0.80 crystal is found to increase linearly along the length of the crystal and then suddenly decreases. The result indicates that the concentration of Na is not stable in the melt which could be the reason for the sudden disruption of the growth.

*3.4 Na-deintercalation process*

$Na_xCoO_2$ ($x$ = 0.30 and 0.60) crystals were prepared by bromination method. To extract sodium, the specimens were cut from the crystal boule with a composition of $Na_{0.75}CoO_2$ and placed in the oxidizing agent $Br_2/CH_3CN$ in a beaker. The temperature was maintained at 30°C and the bromine solution was stirred using magnetic stirrer continuously for about 100 h in order to obtain uniform Na deintercalation. Before and after the extraction process the Na distribution across the crystals were measured by EDX. The result indicated that the average Na content in the resulting crystals after deintercalation reaction is found to have the desired Na content. The deintercalated $x$ = 0.30 crystal could be easily cleaved along the *ab* plane and the cleaved crystal is shown in Fig. 7.

*3.5 Powder XRD studies*

Powder XRD patterns confirmed that all the investigated samples are of single phase. Temperature dependent XRD measurements were performed on crushed single crystals of $Na_{0.75}CoO_2$ using synchrotron radiation [15]. The powders were prepared in closed glass capillaries ($Ø_{inside}$ = 0.3 mm). The powder diffraction patterns were recorded using the OBI detector in the range of 3° ≤ 2$\theta$ ≤ 60° with $\lambda$ = 0.699954 Å. The samples were cooled down to 10 K. The diffraction patterns were taken during heating. Fig. 8 shows Rietveld refinement of a powder diffraction pattern of $Na_{0.75}CoO_2$ at $T$ = 10 K. The unit cell parameters were refined with the program X'Pert Plus [16] using the structure model in space group $P6_3/mmc$ according to Ref. [17]. Fig. 9 shows *a* and *c* lattice parameters variation as a function of temperature for $Na_xCoO_2$ ($x$ = 0.75 and 0.80). In the temperature range of 10 K ≤ $T$ ≤ 300 K the lattice parameters of $x$ = 0.75 and 0.80 increase with increasing temperature. The results confirm the phase purity and crystallanity of the compounds and that there is no structural transition in the investigated temperature range.

*3.6 Magnetic susceptibility*

Static magnetic susceptibility $\chi(T) = M(T)/B$ measurements were carried out on the as grown and Na de-intercalated single crystals. The data show clear dependence of $\chi(T)$ on the Na content and are consistent with the literature values. Note that there are no signs of the common impurity phases such as CoO or $Co_3O_4$ which are known to order antiferromagnetically at around 290 and 35 K, respectively. In literature there are some discrepancies in the $\chi(T)$ data measured on FZ single crystals prepared under slightly different conditions. The observed discrepancies have been attributed to (i) the presence of impurities (ii) compositional inhomogeneity, and (iii) oxygen non-stoichiometry in the crystals. In the present work, we have measured $\chi(T)$ on freshly cleaved $Na_xCoO_2$ samples. The Na content was measured both before and after the $\chi(T)$ measurements. A significant loss of Na was noticed on the crystal surface after exposing the sample to the atmosphere for a few days. However, ICP-OES analysis of the bulk sample confirmed that the average composition is still close to the nominal value with an experimental error of ±0.03.

Fig. 10 shows $\chi(T)$ of $Na_{0.75}CoO_2$, $Na_{0.30}CoO_2$ and $Na_{0.60}CoO_2$ single crystals in a magnetic field of 1 T applied along c-axis in the temperature range of 2–360 K. There is a clear transition ($T_C$) at 22 K in $Na_{0.75}CoO_2$ and $Na_{0.60}CoO_2$ single crystals. $\chi(T)$ drops down below $T_C$ which is attributed to the magnetic phase transition associated with the onset of long-range antiferromagnetic spin density wave-like (SDW) magnetic order with easy magnetic axis along the *c*-axis [18]. The phase transition temperature ($T_C$) of the $x = 0.75$ sample is completely independent of magnetic fields below 5T which is clearly seen in Fig. 11. Above $T_C$ the susceptibility on both the samples obey the Curie-Weiss law and is well fit to the formula $\chi(T) = \chi_0 + C/(T+\Theta)$ in the temperature range 150-300 K where $\chi_0$ is the temperature independent magnetic susceptibility which includes the Pauli paramagnetism, core diamagnetism and van Vleck contribution from Co ions, $C$ the Curie constant and $\Theta$ is the Weiss constant. The fit gives the value of $\chi_0$ as $3.358\times10^{-4}$ emu/mol and $3.4742\times10^{-4}$ emu/mol for $x = 0.75$ and 0.60 samples, respectively. The Weiss constant is approximately -245 K and -336 K for $Na_{0.75}$ and $x = 0.60$ samples, respectively. These negative larger values of $\Theta$ indicate the stronger antiferromagnetic interaction. Thus the Curie-Weiss behavior of $x = 0.75$ and 0.60 samples support the magnetic phase diagram which lie in the Curie-Weiss metal region. The Curie-Weiss behavior is suppressed when we decrease $x$ in $Na_xCoO_2$ from 0.75 to 0.30 and the magnetic susceptibility becomes featureless. The anomaly at 334 K has been reported by Huang et al. [19] to be due to a structural transition for $x = 0.75$ sample (see inset in Fig. 8). This structural transition is shifted to 339 K for the sample with $x = 0.60$.

Besides the two well known characteristic features at 22 K and 334 K, there were two more anomalies such as; (i) a broad maximum around 50 K, (ii) an upward kink at 9 K in $x = 0.75$ sample. But this broad maximum anomaly is not seen in $x = 0.60$ sample. Sakurai et al. [5] have reported similar broad magnetic anomaly at around 50 K ($T_m$) for $Na_{0.78}CoO_2$ sample and have attributed that a heavy Fermi-liquid state develops below this temperature. The $T_m$ value decreases with decreasing Na content and reaches the minimum of 10 K for 0.70 sample. Also Viciu et al. and Okamoto et al. reported that the $x = 0.6$ sample shows only Curie-Weiss behavior but not heavy Fermi liquid state [3, 20]. So the heavy Fermi liquid state may not develop in $x = 0.60$ sample. The anomaly at 9 K has an upturn under low magnetic fields of 0.1 T and 0.2 T whereas a downward kink is seen for the higher fields of 1 T, 2 T and 5 T (see Fig. 11). To know the magnetic nature of $x = 0.75$ sample we have studied the magnetization at different temperatures of 2 K, 10 K, 25 K and 250 K upto the applied magnetic field of 5 T and it is shown in Fig. 12. The figure clearly shows that the magnetization is linear for the temperatures 25 K and 250 K. At 2 K, the data show a non-linear behavior in M vs H which is expected for an antiferromagnet if the field is applied along the magnetic easy axis. The data imply spin orientation at about 2.5-3T indicating a moderate anisotropy of the magnetic moments.

## 4. Conclusions

Thermal analysis of the $Na_xCoO_2$ ($x = 0.75$, 0.80 and 0.90) powder samples suggested that the compounds melt incongruently at 1035 ± 5°C in oxygen atmosphere. However, a significant increase in the Na loss occurred in the samples with higher Na content ($x = 0.80$ and 0.90). Large single crystal $Na_xCoO_2$ ($x = 0.75$ and 0.80) boules have been grown by traveling solvent floating zone (TSFZ) method. Appropriate growth conditions such as rotation of feed and seed rods in the opposite directions (25 rpm), pulling rate (2.5 mm/h) and growth atmosphere (30 ml/min) yielded $Na_{0.75}CoO_2$ single crystals with dimensions of the order of several tens of mm in length and about 6 mm in diameter. Phase purity and quality of the grown crystals have been confirmed by powder XRD and magnetic susceptibility studies. The spin density wave (SDW) transition occurred at 22 K in both $x = 0.60$ and 0.75 samples. However, this transition is found to be sharper in $x = 0.60$ crystal than that of $x = 0.75$ sample which may be due to the development of heavy Fermi liquid around 50 K in the latter case.

**Acknowledgements**

The authors thank the UGC-Govt. of India for financial assistance. S. Paulraj and P. Kanchana acknowledge CSIR-Govt. of India for research fellowship. The authors C. Sekar and

S. Paulraj acknowledge UGC-DAE CSR - Govt. of India for financial assistance. The author R. Klingeler acknowledges DFG via project KL1824/2 for the financial assistance.

74

**Table 1** Decomposition temperature and mass loss of $Na_xCoO_2$ ($x$ = 0.75, 0.80 and 0.90).

| $x$ in $Na_xCoO_2$ | Decomposition temperature | | Mass loss (wt.)% (TG) |
|---|---|---|---|
| | $T$1 | $T$2 | |
| 0.75 Single crystal | - | 1040 | 4.56 |
| 0.75 Polycrystal | - | 1038 | 4.77 |
| 0.8 Polycrystal | 847 | 1039 | 7.52 |
| 0.9 Polycrystal | 850 | 1035 | 7.06 |

**Figures**

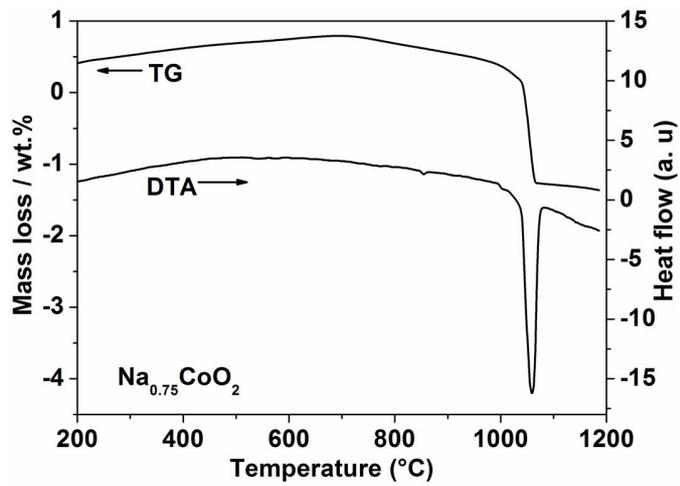

**Fig. 1** TG-DTA profile of $Na_{0.75}CoO_2$ at $p(O_2) = 1$ bar.

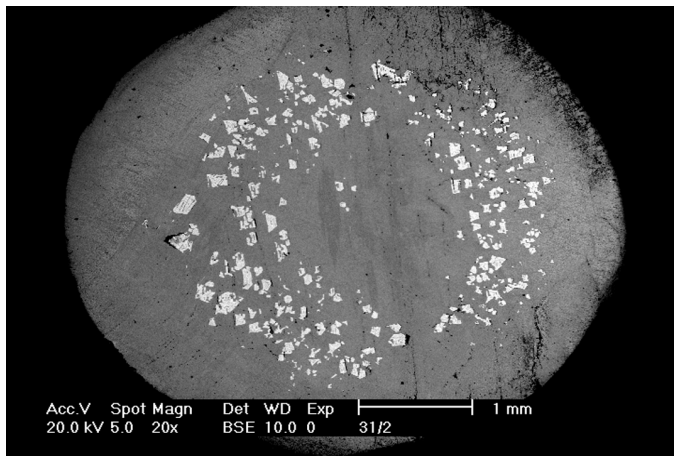

**Fig. 2** Cross sectional view of $Na_{0.75}CoO_2$ crystal boule grown using previously grown crystal as seed rod. White spots are the impurity phases of $CoO_{1-\delta}$.

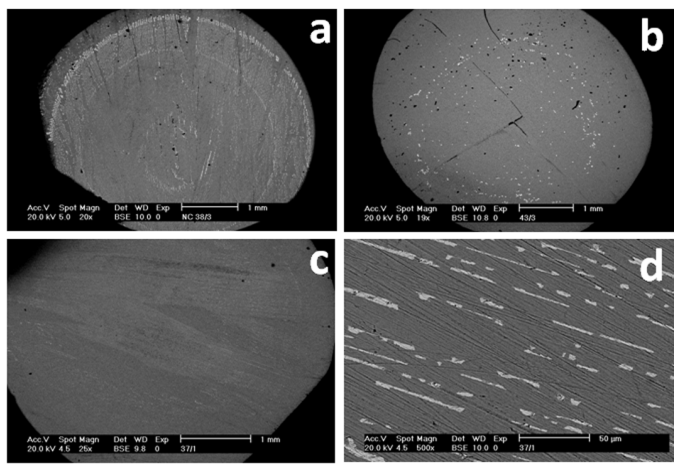

**Fig. 3** SEM pictures of $Na_{0.75}CoO_2$ crystals grown with the rotation rates of (a) 15 rpm, (b) 20 rpm, and (c) 30 rpm under flowing oxygen atmosphere. (d) The stripe-like pattern observed on 30 rpm grown crystal under higher magnification.

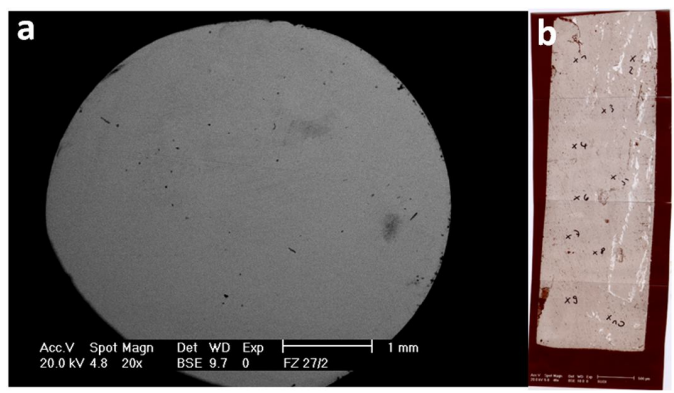

**Fig. 4** SEM pictures of typical $Na_{0.75}CoO_2$ single crystal grown under optimized condition taken along (a) horizontal and (b) vertical cross-section.

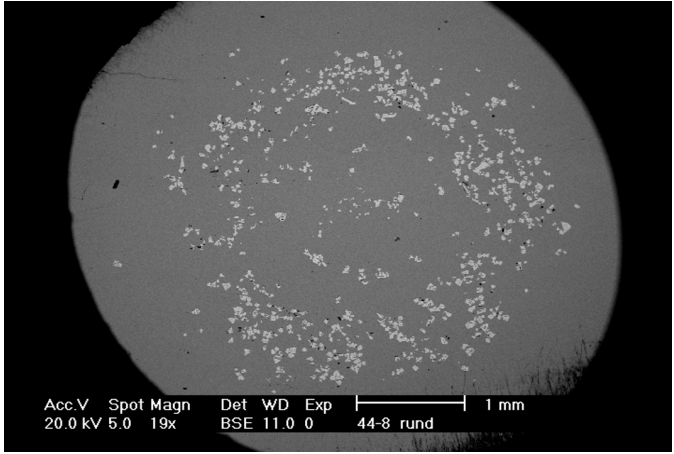

**Fig. 5** SEM picture of grown $Na_{0.80}CoO_2$ single crystal. White spots indicate the impurity phases of $Co_3O_4$.

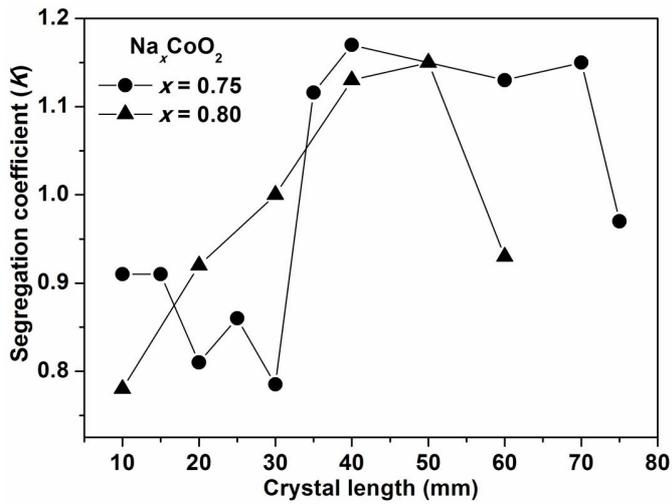

**Fig. 6** Variation of segregation coefficient ($K$) along the longitudinal length of (●) $Na_{0.75}CoO_2$ and (▲) $Na_{0.80}CoO_2$ crystal boules.

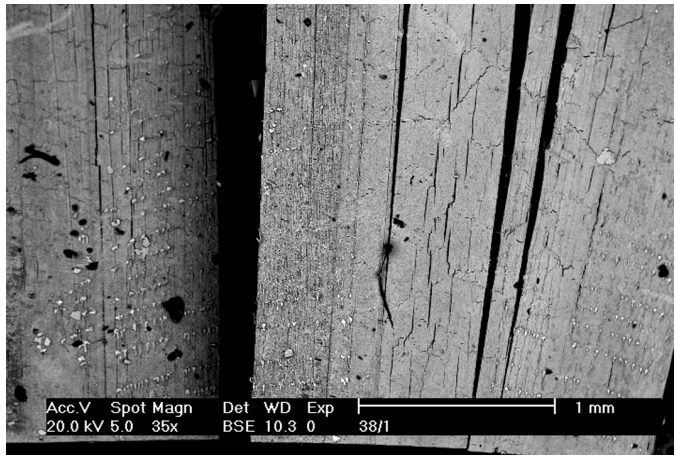

**Fig. 7** Morphology of $Na_{0.3}CoO_2$ single crystal after Na deintercalation from $Na_{0.75}CoO_2$.

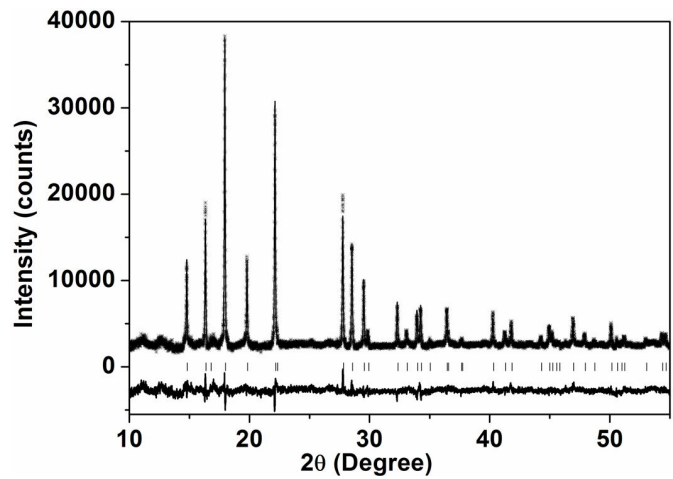

**Fig. 8** Rietveld refinement of a powder diffraction pattern of $Na_{0.75}CoO_2$ at $T = 10$ K.

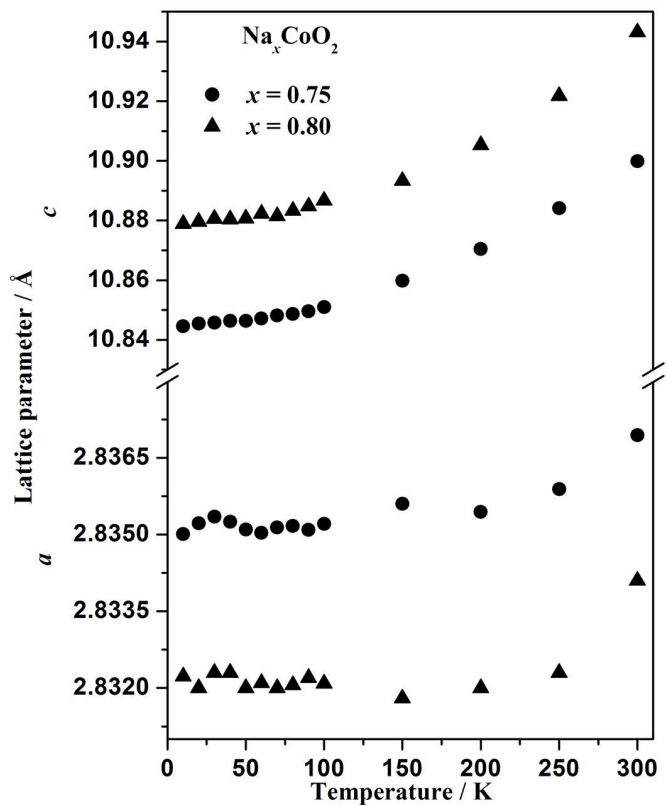

**Fig. 9** Variation of lattice parameters as a function temperature for $Na_xCoO_2$ ($x$ = 0.75 and 0.80).

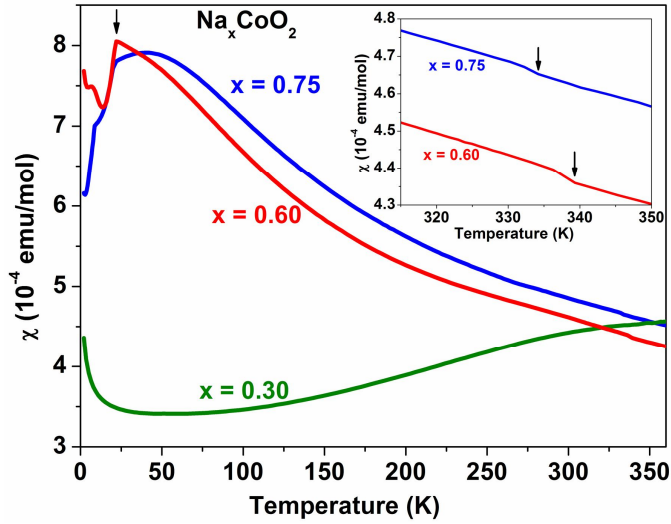

**Fig. 10** Temperature dependence of the DC magnetic susceptibility of $Na_xCoO_2$ ($x$ = 0.30, 0.60 and 0.75) single crystals. Inset: high temperature susceptibility of $x$ = 0.60 and 0.75 samples. Arrows at 339 K and 334 K indicate the structural transition of x = 0.60 and 0.75 samples, respectively.

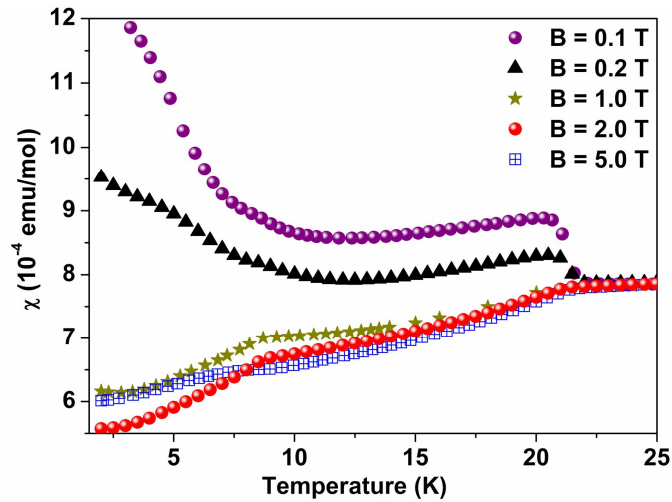

**Fig. 11** Low temperature susceptibility of $Na_{0.75}CoO_2$ single crystal in different applied fields.

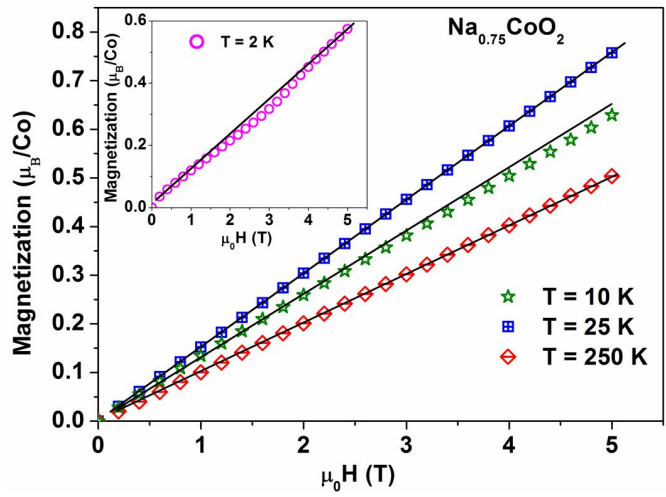

**Fig. 12** Magnetization as a function of magnetic field at different temperatures for $Na_{0.75}CoO_2$ single crystal. Solid lines are guide to the eye. Inset: M-H curve at T= 2 K.